\documentclass{pasj00}

\begin{document}
\SetRunningHead{M.Iye et al.}{Performance and Improvements of the Subaru Telescope}
\Received{2003/10/09}
\Accepted{2004/02/16}

\title{Current Performance and On-Going Improvements of the 8.2 m Subaru Telescope} 

%

%

\author{
Masanori \textsc{Iye}\altaffilmark{1},
Hiroshi \textsc{Karoji}\altaffilmark{2},
Hiroyasu \textsc{Ando}\altaffilmark{1},
Norio \textsc{Kaifu}\altaffilmark{3},   
Keiichi \textsc{Kodaira}\altaffilmark{4}, \\ 
Kentaro \textsc{Aoki}\altaffilmark{2},
Wako \textsc{Aoki}\altaffilmark{1},
Yoshihiro \textsc{Chikada}\altaffilmark{3},
Yoshiyuki \textsc{Doi}\altaffilmark{2},
Noboru \textsc{Ebizuka}\altaffilmark{5},\\
Brian \textsc{Elms}\altaffilmark{2},
Gary \textsc{Fujihara}\altaffilmark{2},      
Hisanori \textsc{Furusawa}\altaffilmark{2},
Tetsuharu \textsc{Fuse}\altaffilmark{2},
Wolfgang \textsc{Gaessler}\altaffilmark{6}, \\
Sumiko \textsc{Harasawa}\altaffilmark{2},
Yutaka \textsc{Hayano}\altaffilmark{1},
Masahiko \textsc{Hayashi}\altaffilmark{2},    
Saeko \textsc{Hayashi}\altaffilmark{2}, 
Shinichi \textsc{Ichikawa}\altaffilmark{1},    \\
Masatoshi \textsc{Imanishi}\altaffilmark{1},
Catherine \textsc{Ishida}\altaffilmark{2}, 
Yukiko \textsc{Kamata}\altaffilmark{7},       
Tomio \textsc{Kanzawa}\altaffilmark{2},      
Nobunari \textsc{Kashikawa}\altaffilmark{1},   \\
Koji \textsc{Kawabata}\altaffilmark{1},
Naoto \textsc{Kobayashi}\altaffilmark{8}, 
Yutaka \textsc{Komiyama}\altaffilmark{2},    
George \textsc{Kosugi}\altaffilmark{2},
Tomio \textsc{Kurakami}\altaffilmark{2},     \\
Michael \textsc{Letawsky}\altaffilmark{2},      
Yoshitaka \textsc{Mikami}\altaffilmark{1}, 
Akihiko \textsc{Miyashita}\altaffilmark{2},
Satoshi \textsc{Miyazaki}\altaffilmark{2}, 
Yoshihiko \textsc{Mizumoto}\altaffilmark{1},\\ 
Junichi \textsc{Morino}\altaffilmark{2},     
Kentaro \textsc{Motohara}\altaffilmark{8}, 
Koji \textsc{Murakawa}\altaffilmark{2},   
Masao \textsc{Nakagiri}\altaffilmark{4}, 
Kyoto \textsc{Nakamura}\altaffilmark{7},  \\
Hidehiko \textsc{Nakaya}\altaffilmark{2},
Kyoji \textsc{Nariai}\altaffilmark{9}, 
Tetsuo \textsc{Nishimura}\altaffilmark{2},   
Kunio \textsc{Noguchi}\altaffilmark{1}, 
Takeshi \textsc{Noguchi}\altaffilmark{10},   \\ 
Junichi \textsc{Noumaru}\altaffilmark{2}, 
Ryusuke \textsc{Ogasawara}\altaffilmark{2}, 
Norio \textsc{Ohshima}\altaffilmark{7},  
Yoichi \textsc{Ohyama}\altaffilmark{2},     
Kiichi \textsc{Okita}\altaffilmark{11},      \\
Koji \textsc{Omata}\altaffilmark{2},          
Masashi \textsc{Otsubo}\altaffilmark{7}, 
Shin \textsc{Oya}\altaffilmark{2},       
Robert \textsc{Potter}\altaffilmark{2},  
Yoshihiko \textsc{Saito}\altaffilmark{1},  \\
Toshiyuki \textsc{Sasaki}\altaffilmark{2},
Shuji \textsc{Sato}\altaffilmark{12}, 
Dennis \textsc{Scarla}\altaffilmark{2},    
Kiaina \textsc{Schubert}\altaffilmark{2},
Kazuhiro \textsc{Sekiguchi}\altaffilmark{2}, \\
Maki \textsc{Sekiguchi}\altaffilmark{13}, 
Ian \textsc{Shelton}\altaffilmark{14}, 
Chris \textsc{Simpson}\altaffilmark{15},      
Hiroshi \textsc{Suto}\altaffilmark{2}, 
Akito \textsc{Tajitsu}\altaffilmark{2},        \\
Hideki \textsc{Takami}\altaffilmark{2},    
Tadafumi \textsc{Takata}\altaffilmark{2}, 
Naruhisa \textsc{Takato}\altaffilmark{2},   
Richard \textsc{Tamae}\altaffilmark{2}, 
Motohide \textsc{Tamura}\altaffilmark{1},  \\ 
Wataru \textsc{Tanaka}\altaffilmark{16},
Hiroshi \textsc{Terada}\altaffilmark{2},     
Yasuo \textsc{Torii}\altaffilmark{1},       
Fumihiko \textsc{Uraguchi}\altaffilmark{2}, 
Tomonori \textsc{Usuda}\altaffilmark{2},  \\
Mark \textsc{Weber} \altaffilmark{2},            
Tom \textsc{Winegar}\altaffilmark{2}, 
Masafumi \textsc{Yagi}\altaffilmark{1},                    
Toru \textsc{Yamada} \altaffilmark{1},
Takuya \textsc{Yamashita}\altaffilmark{2},  \\ 
Yasumasa \textsc{Yamashita}\altaffilmark{17},
Naoki \textsc{Yasuda}\altaffilmark{1},
Michitoshi \textsc{Yoshida}\altaffilmark{11}, 
and
Masami \textsc{Yutani}\altaffilmark{1}
}

\altaffiltext{1}{Optical and Infrared Astronomy Division, National Astronomical 
Observatory, Mitaka, Tokyo 181-8588}
\email{iye@optik.mtk.nao.ac.jp}
\altaffiltext{2}{Subaru Telescope, National Astronomical Observatory, 650 North 
A'oh\={o}k\={u} Place, Hilo, HI 96720, USA} 
\altaffiltext{3}{National Astronomical Observatory, Mitaka, Tokyo 181-8588}
\altaffiltext{4}{Department of Astronomy, Graduate University for Advanced Studies, 
Hayama, Kanagawa 240-0193}
\altaffiltext{5}{The Institute of Physical and Chemical Research (RIKEN), Wako, 
Saitama 351-0198}
\altaffiltext{6}{Max-Plank f\"{u}r Astronomie, K\"{o}nigstuhl 17, 69117 Heidelberg, 
Germany}
\altaffiltext{7}{Advanced Technology Center, National Astronomical Observatory, 
Mitaka, Tokyo 181-8588}
\altaffiltext{8}{Institute of Astronomy, School of Science, University of Tokyo, 
Mitaka, Tokyo 181-8588}
\altaffiltext{9}{Department of Physics, Meisei University, Hino, Tokyo 191-8506}
\altaffiltext{10}{290-2 Izawa, Misaki-cho, Isumi-gun, Chiba 299-4615}
\altaffiltext{11}{Okayama Astrophysical Observatory, National Astronomical 
Observatory, Asakuchi-gun, Okayama 719-0232}
\altaffiltext{12}{Department of Astrophysics, Nagoya University, Chikusa-ku, Nagoya 
464-8602}
\altaffiltext{13}{5-5-22-201 Sekimae, Musashino-shi, Tokyo 180-0014}
\altaffiltext{14}{David Dunlap Observatory, 123 Hillsview Drive, Richmond Hill, 
Ontario, L4C 1T3 Canada}
\altaffiltext{15}{Department of Physics, Durham University, South Road, Durham. DH1 
3LE, UK}
\altaffiltext{16}{4-4-23 Takatanobaba, Shinjuku-ku, Tokyo 160-0075}
\altaffiltext{17}{1-14-12 Asahigaoka, Hino, Tokyo 191-0065}


\maketitle

\begin{abstract}
An overview of the current status of the 8.2~m Subaru Telescope constructed and 
operated at Mauna Kea, Hawaii, by the National Astronomical Observatory of Japan is 
presented. The basic design concept and the verified performance of the telescope 
system are described. Also given are the status of the instrument package offered 
to the astronomical community, the status of operation, and some of the future plans. 
The status of the telescope reported in a number of SPIE papers as of the summer of 
2002 are incorporated with some updates included as of 2004 February.  However, 
readers are encouraged to check the most updated status of the telescope through the 
home page, http://subarutelescope.org/index.html, and/or the direct contact with the 
observatory staff.
\end{abstract}

\section{History of the Project Planning and Construction}
Conceptual planning for the 8.2 m (originally 7.5~m) Subaru Telescope began in the 
summer of 1984 when an engineering working group was established by the Tokyo 
Astronomical Observatory of the University of Tokyo with full support from the optical 
and infrared astronomy community.  In 1985 March,  the Committee for Astronomy of the 
Science Council of Japan recommended the construction of the Japan National Large 
Telescope (JNLT) \citep{kod86}, the original name of the project that subsequently 
produced the Subaru Telescope, as the top priority amongst all other projects planned 
by the astronomical community. A Memorandum of Understanding between the Tokyo 
Astronomical Observatory and the University of Hawaii to sublease part of the summit 
area of Mauna Kea in Hawaii as a reserved site for the JNLT was signed in the summer 
of 1986. The first report summarizing the results of detailed studies was compiled 
in the ``White Book" in 1986 September. In 1988 July, the Tokyo Astronomical Observatory 
was reorganized  as an inter-university institute, the National Astronomical 
Observatory (NAOJ), to promote large-scale enterprises such as the JNLT project on 
a national basis \citep{kod89}. 

In 1989, active optics experiments were carried out for proof of concept \citep{iye89, 
nog89, iye90}, and the effect of mirror seeing was quantitatively evaluated 
\citep{iye91}. At the same time, site tests were performed at Mauna Kea \citep{and89}, 
and wind and water tunnel tests were carried out to identify the optimum shape of 
the enclosure \citep{and91}.

A second advanced report, the ``Blue Book", was compiled in 1989, in which most of 
the baseline features of the JNLT were presented, and the project was announced later 
that year at an international conference \citep{kod90}. Acknowledging the dimensions 
and capabilities of the Keck Telescope and the Very Large Telescope of the European 
Southern Observatory, which began construction during the course of the 6-year 
feasibility study for JNLT, the target diameter of the primary mirror was expanded 
from 7.5 m to 8.2 m, without increase in the total budget.  

Following the recommendations of the Science Council of Japan, the project was fully 
approved and construction of JNLT began in 1991 April \citep{kog91}. A call for 
proposals for renaming JNLT ended up with its new name, Subaru Telescope, after 
the old and familiar Japanese name for the visible star cluster Pleiades \citep{kod92}. 
Progress of the project and the instrumentation plan was critically reviewed at another 
international conference held in Tokyo in 1994 \citep{iye95}. The Subaru Telescope 
base facility  was established at the University of Hawaii campus in Hilo, Hawaii, 
in 1997 April, and more than 20 NAOJ staff relocated from the Mitaka campus to begin 
work on the telescope in collaboration with local staff recruited through the Research 
Corporation of the University of Hawaii. The progress of construction was reported 
on several occasions \citep{iye97, kai98}, and the engineering First Light of the 
Subaru Telescope was achieved on 1998 December 24. Reports on First Light observations 
were published in 2000 \citep{kai00, iye00}. An inauguration ceremony was held in 
1999 September. The 9-year construction project was completed in 2000 March, and the 
telescope has been made available to both Japanese and the international community 
of astronomers since 2000 December.

   \begin{figure}
   \begin{center}
   \end{center}
   \caption{Concept of the active support system for the primary mirror. \label{kanzf1}}
   \end{figure} 

\section{Telescope Structure}
\subsection{Optics}
The Subaru Telescope has an 8.2 m \it{F}\rm/1.83 primary mirror with a focal length of 15 
m. Three secondary mirrors are installed, each comprising a modified Ritchey--Cretien 
system at the \it{F}\rm/12.2 Cassegrain or \it{F}\rm/12.6 Nasmyth foci. The basic optical parameters 
are compiled in table 1. 

The wide-field prime focus is the most notable feature of the Subaru Telescope, and 
is a unique capability among 8 -- 10 m telescopes. An ambitious wide-field prime-focus 
corrector with a new type of atmospheric dispersion compensator \citep{nar94} allows 
for a 30\arcmin field of view at \it{F}\rm/2 with an optical image quality of better than 
0\farcs23, right up to the edge of the field.

The Cassegrain focus offers a 6\arcmin field of view in combination with either the 
optical or infrared secondary mirror, depending on the instrument to be used. The 
optical Nasmyth focus utilizes a dedicated secondary mirror to cover a 4\arcmin field 
of view, and the infrared Cassegrain secondary can be used for observations at the infrared 
Nasmyth focus by moving the infrared secondary mirror along the optical axis and 
applying a force correction on the primary mirror shape. Although the coating and pupil 
size are different, the infrared secondary can be used for optical Nasmyth observations, 
and the optical secondary can be used for infrared Nasmyth observations, if necessary.

The infrared secondary mirror has chopping and tip-tilt correction capabilities, 
which are useful for near-infrared observations and indispensable for mid-infrared 
observations. 

\begin{longtable}{lcccc}
  \caption{Parameters of optical components of the Subaru 
Telescope.}\label{tab:optics}
  \hline\hline
  Mirror & Diameter & Radius of & Aspheric & Distance to \\
       &   & curvature & constant & previous surface \\
       & (mm) & (mm) &    & (mm) \\
  \hline
  M1(Primary) & 8200 & -30000 & -1.008350515 &  -- \\
  M2(Opt/Cass) & 1330 & -5524.297 & -1.917322232 & -12652.174 \\
  M2(Opt/Nas) & 1400 & -5877.420 & -1.865055214 & -12484.300 \\
  M2(IR/Cass) & 1265 & -5524.297 & -1.917322232 & -12652.174 \\
  \hline
\endlastfoot
\end{longtable}

\begin{longtable}{lccccccc}
  \caption{Parameters of seven focal configurations.}\label{tab:optics}
  \hline\hline
Focus & Primary & Cass. & Nasmyth & Nasmyth & Cass. & Nasmyth & Nasmyth \\
wavelength & optical & optical & optical & optical & infrared & infrared & infrared 
\\
  \hline
Image  derotation & Instr. Rot. & Instr. Rot. & Ima. Rot. & None & Instr.Rot. & Ima. 
Rot. & None \\
Ritchey--Cretien & NA & Yes & Yes & No & Yes & No & No \\
Focal length (mm) & 15000 & 100000 & 104207.0 & 102244.5 & 100000 & 108512.4 
& 110605.2 \\
Pupil diameter (mm) & 8200 & 8200 & 8200 & 8200 & 8081.9 & 7971.6 & 7947.5 \\
Back focus (mm)&  NA & 3000 & 4992.6 & 4600 & 3000 & 4600 & 4992.6 \\
Effective \it{F} \rm ratio & 1.83 & 12.195 & 12.708 & 12.469 & 12.373 & 13.612 & 13.917 \\
Field of view (arcmin) & 30 & 6 & 3.5 & 6 & 6 & 3.5 & 6 \\
Field curvature (mm) & NA & 5630.3 & 6108.5 & 6106.6 & 5630.3 & 5635.2 & 5636.3 \\
Plate scale &  13.75099 & 2.062658 & 1.979385 & 2.017378 & 2.062658 & 1.900850 & 
1.864884 \\
  \hline
\endlastfoot
\end{longtable}

\begin{figure}
  \begin{center}
  \end{center}
  \caption{Structure of the mirror support actuator. \label{actuator}}
\end{figure}

\subsection{Active Mirror Support}
Right from the beginning of the feasibility study, which began in 1984, fabrication 
of a light-weight mirror and the associated support system was identified as a critical 
issue. Finite element method (FEM) analyses were conducted to compare the deformation 
of the primary mirrors fabricated from honeycomb borosilicate glass and thin meniscus 
zero-expansion glass \citep{wat87}. It became obvious that a conventional passive 
support would not provide the target image quality for such a large light-weight 
mirror, and a real-time force-controlled mirror support system to adjust for any surface 
shape error was considered to be mandatory. In light of this, borosilicate glass, which 
has a thermal expansion coefficient more than 100-times higher than that of thin meniscus 
glass, became less desirable in terms of the stringent thermal control required for 
the massive primary mirror. The study group also judged that force control, though 
not an easy approach, is more tractable than temperature control. As the approach 
to use and control segmented mirrors, rather than a monolithic mirror, did not appear 
to be feasible at the time of decision making, monolithic, thin meniscus zero-expansion 
glass was adopted for the primary mirror. Following FEM studies of the support system, 
a support system consisting of 261 actuators distributed in 8 rings was adopted to 
support the mirror. The mechanism of support, itself, is also unusual in that instead 
of supporting the mirror using supports mounted orthogonally on the back surface, 
which is an edge-moment support that induces sigmoidal deformation of the tilted mirror, 
the actuators are inserted into ``pockets" bored into the back of the blank at the 
261 actuator locations (cf. figure~\ref{kanzf1}). This configuration effectively 
supports the glass mirror at local centers of gravity, avoiding any sigmoidal 
deformation and significantly reducing the overall bending. The decision to proceed with 
this challenging design was difficult, because the thinness of the mirror blank and the 
many pockets that needed to be drilled from the back greatly increased the potential 
risk of breaking the glass. However, the study team decided to control the risk by 
carefully designing the production processes and the supporting structure considering 
the higher imaging performance expected from this concept.

The actuator mechanism for this mirror support system is shown in 
figure~\ref{actuator}. The support force is applied by a spring on a ball screw, and 
the force is measured by a precise force sensor based on a tuning-fork quartz oscillator. 
The sensor was designed to achieve a $10^{-5}$ resolution and a dynamic range of 
0--1500~N. The temperature coefficient is measured and calibrated for each actuator. 
In this configuration, both the axial and lateral support forces are applied at the 
local center of gravity of the blank.  Lateral support is accomplished passively with 
a counterweight.

\begin{figure}
  \begin{center}
  \end{center}
  \caption{Surface error map for the 8.2~m primary mirror as shown in 20nm interval 
contour map.\label{14nmrms}}
\end{figure}

\subsection{Fabrication of the Primary Mirror}
The primary mirror blank, 8.3~m in diameter and 30~cm in thickness, was manufactured 
using Ultra Low Expansion (ULE) glass by Corning Incorporated, Canton, New York. To 
ensure the quality of the final mirror, the monolithic blank was fabricated from 44 
hexagonal ULE blocks with thermal expansion coefficients in the range ${\pm 5}$ ppb, 
manufactured over the first two years. Some of the hexagonal blocks were cut further 
into two or three pieces to give a total of 55 pieces to form a circular disk, and 
were then arranged and thermally fused in a heating furnace to form the 8.3~m monolithic 
blank. The pieces were arranged according to annealing simulations based on their 
expansion coefficients so as to minimize the residual surface error and in 
consideration of active correction for thermal deformation \citep{mik92}. The optimum 
solution, in fact, provided a residual surface error that was an order of magnitude 
smaller than that given by the random solutions. Thermal fusion of the pieces into 
a monolithic structure was completed early in 1994. After grinding the front surface, 
the blank was turned over to grind the back surface. Finally, the blank was slumped 
over a spherical mold in a heating furnace to an approximate meniscus shape so as to 
reduce the amount of grinding work required \citep{sas94}. 

The thin meniscus mirror blank was then transported in the summer of 1994 to Contraves 
Brashear Corporation near Pittsburgh for polishing. The first operation was to bore 
261 pockets on the back surface of the blank for inserting the invar sleeves to support 
the mirror.  Each pocket was 15 cm deep by 15 cm in diameter. The drilled surface 
was then polished and acid-etched to remove any remaining micro cracks on the surface. 
The mirror blank was again turned over in the autumn of 1995 for processing of the 
front surface. Grinding and polishing was carried out in a 36 m-high underground shaft 
constructed at the large optics facility of Contraves Brashear Corporation, Wampum, 
near Pittsburgh. This stable underground facility has advantages over the 
conventional tower facility, avoiding vibrations induced by wind load and traffic, 
as well as diurnal tilt of the shaft induced by the thermal load of the sunshine to 
the tower.

The primary mirror was set on a rotating polishing bed with 261 hydraulic supports 
distributed identically to the actual active support system. The polishing procedure 
was executed using a computer-controlled polishing machine, ensuring careful 
maneuvering of the work functions. The progress of polishing was monitored by various 
independent measurements, including profilometer, infrared interferometer, optical 
interferometer, and penta prism measurements. 

The finished mirror has a physical diameter of 8.3~m, a thickness of 20~cm and a weight of 
22.8 tonnes. The diameter of the effective reflecting surface is 8.2~m, and the focal 
length is 15~m. The final surface error after compensating for the first 21 modes 
of deformation by the active optics system was demonstrated to be 14 nm rms on 1998 August 
28. The surface error map is shown in figure~\ref{14nmrms}.

\begin{figure}
  \begin{center}
  \end{center}
  \caption{Schematic of telescope structure. \label{telefoci}}
\end{figure}

\subsection{Telescope Structure}
On the alt-azimuth mount structure, the entire telescope is 22.2~m in height and 27.2~m 
wide, with a moving assembly weighing 555 tonnes (figure~\ref{telefoci}). The 
telescope can be driven at a maximum slew rate of $0.^\circ5 /s$, and the cable wrap 
allows for rotation of $270^\circ$ in each direction from the south. The top ring and 
the mirror cell are supported from the center section by trusses to ensure minimum 
flexure and to maintain the optical alignment between the primary and secondary mirrors. 
The secondary mirror unit is suspended by a torsion-resistant spider with a special 
locking mechanism to secure the secondary mirror units or primary focus unit into 
the inner ring supported by the spider \citep{miy94}.

The tertiary mirror tower houses two tertiary mirrors, one for the optical Nasmyth 
and the other for the infrared Nasmyth foci. These tertiary mirrors can be inserted 
and removed in about 12 minutes. The primary mirror cover consists of 23 separate 
panels suspended by wires from the center section and 4 side covers that can be opened 
or closed in about 7 minutes when the telescope is pointing to the zenith.

The mirror cell houses 261 actuators to support the primary mirror. There are three 
additional fixed point devices with a mechanical fuse mechanism to avoid the accumulation 
of excess force at these fixed points. The mechanical fuses are triggered (released) 
to protect the mirror when the force applied at the fixed points exceeds a threshold 
value, 1.5 tonnes, as may occur during an earthquake, for example. In the initial 
phase of operation, the metal pad fixed to the primary mirror for this purpose became 
unstuck on a few occasions, attributed to imperfect adhesion. New pads were designed 
and replaced in the summer of 2001 to ensure functionality in the event of larger 
forces at the adhering surface. The mechanical fuse was also readjusted to function 
at a lower level of 0.5 tonnes force to ensure the safety of the primary mirror.

An acquisition and guide system consisting of a Shack--Hartmann camera, a guider unit, 
and a calibration lamp unit is installed at each operating focus. The Cassegrain unit 
also houses the adaptive optics system, located in front of the acquisition and guiding 
system. A hydrostatic bearing system, driven by a direct-drive linear motor, ensures 
very smooth pointing and tracking operation.  

Actual fabrication and pre-assembly of the telescope structure was completed in Osaka 
in 1996 \citep{nog98}, and the disassembled parts were shipped to Hawaii for 
reassembling at the summit, which began in 1996 October. Installation of the telescope 
structure in the enclosure was completed in 1998 without incident. The actual 
performance of the telescope is discussed later.

\begin{figure}
  \begin{center}
  \end{center}
  \caption{Water tunnel test showing the stream lines around the 
enclosure.\label{DOMESHAP}}
\end{figure}

\begin{figure}
  \begin{center}
  \end{center}
  \caption{Subaru enclosure vertical layout. \label{domedra}}
\end{figure}

\begin{figure}
  \begin{center}
  \end{center}
  \caption{Aerial view of the Subaru enclosure. \label{domeheli}}
\end{figure}

\begin{figure}
  \begin{center}
  \end{center}
  \caption{Top unit handling system. \label{kuraf5}}
\end{figure}

\subsection{Enclosure and Support Facilities}
A series of studies was conducted to compare the aerodynamic performance of the 
enclosure with the traditional hemi-spherical shape, cylindrical shape and others 
\citep{mik89, and91, mik94}. It was found that a cylindrical enclosure is superior 
to a hemi-spherical dome in terms of preventing the turbulent surface boundary layer 
from advancing up along the enclosure and deteriorating the observation conditions 
(figure~\ref{DOMESHAP}). In order to ensure smooth flow of flushing air to carry away 
any residual warm air near the telescope environment, two huge walls, named ``The Great 
Wall", were constructed, along with a number of ventilators and louvers to control 
the air flow inside the enclosure.

The Subaru site was chosen at a latitude of $19^\circ 49\arcmin43\arcsec$N, longitude of 
$155^\circ 28\arcmin50\arcsec$W, and altitude of 4139 m \citep{and89}. The construction of the lower 
part of the enclosure building began on 1992 July 6. The ground cinder soil of the 
site was first stiffened by adding huge amount of cement and water. A robust concrete 
foundation was laid to support the 2000 tonnes of the enclosure, and the telescope 
pier, 29 m in diameter and 14 m high, was then erected, followed by the construction 
of the steel reinforced concrete building of the lower part of the enclosure. 

Construction of the upper part of the enclosure, which corotates with the telescope, 
was initiated in the summer of 1994. The entire structure was built on the azimuth 
rail at the top of the lower building. The enclosure has five floors; the observation 
floor, Nasmyth floor, tertiary floor, top unit floor, and top crane floor 
(figure~\ref{domedra}).  The outer wall is covered with aluminum panels, and the 
structure has an overall height of 43 m and a base diameter of 40 m 
(figure~\ref{domeheli}).  The aluminum outer panel reflects the blue sky and appears 
less conspicuous than conventional white painted domes.  Also, the aluminum panel 
has lower emissivity than the white painted domes and does not cool down excessively 
during the night. It is beneficial for reducing the temperature difference between 
the enclosure and the ambient air.

 All the secondary mirror units and the prime focus unit are housed in a carousel 
on the top-unit floor. Exchanging these top units is achieved by manipulating the 
top unit handling system \citep{kura02}, and takes 3 to 6 hours (figure~\ref{kuraf5}).

\begin{figure}
  \begin{center}
  \end{center}
  \caption{Layout of the coating facilities and mirror handling equipment on the ground 
floor of the enclosure.\label{yutanif1}}
\end{figure}

\subsection{Mirror Coating Facilities}
The mirror handling facility, mirror washing facility, mirror cell trolley, and mirror 
coating chamber are located on the basement floor of the enclosure 
(figure~\ref{yutanif1}). Realuminizing the primary mirror requires the telescope to 
be non-operational for a minimum of two weeks. The first step in the realuminization 
operation is to lift the mirror cell trolley from the ground floor to the observation 
floor through the 10 m mirror hatch using the 80 tonne top crane. The mirror cell 
trolley then accepts the mirror cell from the telescope. The top crane lowers the 
mirror cell with the primary mirror down to the ground floor , where the primary mirror 
is detached from the mirror cell using a hoist. After being transferred to the lower cell of 
the mirror vacuum chamber, the mirror is moved in underneath the mirror washing 
facility. The mirror is washed and then sprayed with hydrochloric acid to remove any 
existing aluminum film. 
All the chemicals and waste water used to this operation is gathered and disposed 
in accordance with the local regulation. Several steps follow washing and drying the 
surface. Finally, the lower cell is moved to underneath the upper cell of the vacuum 
coating facility \citep{yuta02}.

The vacuum-coating chamber contains aluminum filaments for the evaporative coating of 
aluminum \citep{sae98}.  Due to power limitations, aluminum deposition is performed 
by firing the filaments in three grouped zones. The measured reflectivity of the 
aluminized witness sample verifies that the reflectivity exceeds ${91\%}$ at 500 nm 
\citep{kama02}.

\begin{figure*}
  \begin{center}
  \end{center}
  \caption{Subaru computer network linking the summit, Hilo base, and Mitaka. 
\label{HILONET}}
\end{figure*}

\begin{figure*}
  \begin{center}
  \end{center}
  \caption{Subaru observation control system architecture. (Right) Telescope control 
system, (left) observation control system, (upper) workstations used as user 
interfaces, (middle) control servers, and (bottom) hardware control systems. 
\label{SOSSbloc}}
\end{figure*}

\subsection{Control System and Data Archive System}
Figure~\ref{HILONET} shows the configuration of the computer systems at the summit, 
Hilo base facility, and the Mitaka campus. A dedicated optical-fiber link connects 
the summit to the Hilo base facility, and the Hilo base facility to the Mitaka campus 
to ensure swift transfer and backup of observation data.  Remote monitoring and remote 
operation of the telescope has been tested, and a range of observations can be performed 
by linking the summit either to Hilo base or to Mitaka.

Figure~\ref{SOSSbloc} illustrates the overall architecture of the observation 
control system for the Subaru Telescope. The kernel of the telescope control (TSC) 
system is a TSC server linked to three user interfacing workstations, TWS1--3. This 
system controls the mid-level processors, MLP1--3. MLP1 controls all of the driving 
function of the telescope, enclosure, instrument rotators, image rotators, 
atmospheric dispersion correctors, and other peripheral optics devices. MLP2 handles 
the active support system of the primary mirror, and MLP3 monitors and controls the 
telescope environment. Actual control of each component is carried out by dedicated 
board computers, called local control units (LCUs). All of these computers are linked 
by local-area networks \citep{oga98, sas98, kos98, nou98, tana98}. Control of 
observations is supervised by the supervisor workstation (OBS). Observers can send 
commands from observation operation terminals OWS1--3 to OBS. The observational 
instruments are controlled by the instrument control computer (OBCP), which is 
interfaced to OBS and the data-acquisition system (OBC).

   \begin{figure*}
   \begin{center}
   \end{center}
   \caption{Distributed data archive systems for the Subaru Telescope. \label{taktf2}}
   \end{figure*} 

A huge data archive system with 600 TB of tape storage capacity and a supercomputer 
are installed to cater for high-load data analyses, simulations, and weather forecast 
of Mauna Kea \citep{oga02a, oga02b}. The entire software system consists of three 
packages; the Subaru Observation Software System (SOSS) \citep{sas98}, the Subaru 
Telescope data Archive System (STARS) \citep{oga98, takt00}, and the Distributed 
Analysis System Hierarchy (DASH) \citep{miz00, yagi02, oga02c}.  
Figure~\ref{taktf2} shows the configuration of these systems at the Hawaii and Mitaka 
campuses.

\section{Common-Use Instrument Suite}
The Subaru observational instruments plan and its status have been reported in several 
papers \citep{iye95, iyey00, yama02}.  The entire Subaru instrument suite consists 
of 7 instruments: 1) Subaru prime-focus camera (Suprime-Cam), 2) Faint Object Camera 
And Spectrograph (FOCAS), 3) High-Dispersion Spectrograph (HDS),  4) InfraRed Camera 
and Spectrograph (IRCS), 5) OH airglow Suppression spectrograph (OHS),  6) COoled 
Mid-Infrared Camera and Spectrograph (COMICS), and 7) Coronagraphic Imager with 
Adaptive Optics (CIAO). These instruments are deployed at the prime focus 
(Suprime-Cam), optical Nasmyth focus (HDS), infrared Nasmyth focus (OHS), and 
Cassegrain focus (FOCAS, IRCS, COMICS, and CIAO) as shown in figure~\ref{JNLTSTRC}.

Figure~\ref{instpara} shows the region in the wavelength--spectral resolution plane 
covered by these instruments. Table 3 is an updated summary of the key parameters 
of these instruments.  All of the instruments employ closed-cycle refrigerators to cool 
the detector and/or the instrument, and liquid nitrogen is not used for the CCDs.


\begin{figure}
  \begin{center}
  \end{center}
  \caption{Available open-use instruments for the Subaru Telescope.\label{JNLTSTRC}}
\end{figure}

\begin{figure}
  \begin{center}
  \end{center}
  \caption{Observational capability of open-use instruments.\label{instpara}}
\end{figure}


\begin{longtable}{lllllll}
  \caption{Parameters of common-use instruments.}\label{tab:instrum}
  \hline\hline
  Instrument & Focus & Modes of & Spectral & Spectral & FOV & Pixel \\
          &     &  observation & coverage & resolution &  & scale \\
  \hline
  Suprime-Cam & Prime & Im & \it{B,V,R,I,z}, \rm NBF & 10--100 & $34\arcmin \times 27\arcmin$ & 
$0\farcs2$ \\
  FOCAS & Cass & Im/MOS & 0.36 -- 0.90 $\mu$m & 10 -- 2000 & $6\arcmin\phi$ & 
$0\farcs1$ \\
  HDS & Nas & Sp & 0.3 -- 1.0$\mu$m & 100000 & $10\arcsec \times 0\farcs4$ & 
$0\farcs13$ \\
  IRCS & Cass & Im/Sp & 1.0 -- 5.4 $\mu$m & 10 -- 20000 & $1\arcmin\times1\arcmin$ & 
$0\farcs023/0\farcs075$  \\
  OHS/CISCO & Nas & Im/Sp & \it{J,H,K} \rm & 10 -- 800 & $2\arcmin\times2\arcmin$ & $0\farcs12$ \\
  CIAO & Cass & Im/Sp & \it{z,J,H,K,L',M'} \rm & 10 -- 600 & $22\arcsec\times22\arcsec$ & 
$0\farcs012/0\farcs022$ \\
  COMICS & Cass & Im/Sp & \it{N,Q} \rm & 10 --10000 & $42\arcsec\times31\arcsec$ & 
$0\farcs13/0\farcs165$ \\
  \hline
\endlastfoot
\end{longtable}

\subsection{Suprime-Cam}
Suprime-Cam, covering a ${34\arcmin \times 27\arcmin}$ \rm field of view with an unvignetted 
area of ${30\arcmin}$ \rm diameter, by ten 4k ${\times}$ \rm 2k CCDs, is the most frequently 
used Subaru Telescope instrument. The wide field, high quality, prime focus corrector 
provides superb image quality as good as $0\farcs3$ even near the edge of the field under 
a good seeing condition. The median seeing size in the ${I_{\rm C}}$\rm-band is 0\farcs61 and the mean 
of the median seeing size in all the bands is 0\farcs69.  The stellar images are round 
and the mean ellipticity of the point sources is $2.3\pm1.5$\%. Standard sets of 
broad band filters ${B, V, R_{\rm C}, I_{\rm C}, g', r', i'}$, \rm and \it{z'} \rm and a number of narrow band 
filters are available with some restriction on usage. Suprime-Cam has an automatic 
filter exchanger that can hold up to 10 filters.  Upgrading the readout electronics 
to Messia V enabled fast readout within 53 sec to increase the observing efficiency 
\citep{miya98, komi02, miya02}. 

\subsection{FOCAS}
The FOCAS is a Cassegrain instrument with high-throughput all-refractive optics 
optimized for a wavelength region of 365--900 nm. \rm  The instrument has an unvignetted 
field of view of 6\arcmin at 0\farcs1/pixel sampling with two buttable 4k ${\times}$ \rm
2k pixel CCDs mounted with a butting separation of less than ${100 \mu}$m.  \rm 
The FOCAS has a direct imaging mode, a long-slit spectrographic mode, a multi-slit 
spectrographic mode, and a polarimetric imaging and spectroscopic mode.  A spectral 
resolution of ${R = 250 - 2000}$ is available for a slit width of 0\farcs4 by selecting 
one of the 4 grisms. High-dispersion grisms and an echelle grism are under verification 
tests and will become available soon \citep{kas00, yos00, kas02, kawa02, sait02}.

\subsection{HDS}
The HDS is a Nasmyth echelle spectrograph with a quasi-Littrow configuration 
\citep{nog02}. A catadioptric camera with a triplet corrector is used to allow covering 
a wide wavelength region of ${0.3 - 2.3 \mu}$m.  However, only optical CCDs are 
provided as the detector. An echelle grating with 31.6 gr mm$^{-1}$ \rm blazed at 
${71.^{\circ}5}$ is used as the primary disperser. Two cross-disperser gratings, 
optimized for the blue and red regions, are also installed, and one is selected to 
obtain the optimum order separations on the CCDs. The HDS is designed to provide a spectral 
resolution of $R$ \rm = 100000 with a 0\farcs4-wide slit, which projects onto ${45 \mu}$m, or 3 pixels, of the CCD.

\begin{figure*}
  \begin{center}
  \end{center}
  \caption{Block diagram of the active support system for the 8.2 m primary mirror. 
\label{takaf1}}
\end{figure*}

\subsection{IRCS}
The IRCS is a common-use instrument for near-infrared (1-5 ${\mu}$m) imaging 
and spectroscopy at diffraction-limited spatial resolution. The instrument was 
constructed at the Institute for Astronomy of the University of Hawaii. An echelle 
spectroscopy mode is provided in addition to the imaging mode and the grism spectroscopy 
mode \citep{kob00, tera02}.

\subsection{OHS}
The OHS is an original instrument designed and constructed at Kyoto University 
\citep{mai93}. The instrument has a first-stage medium resolution $(R \sim 
5000)$ spectrograph that produces an infrared spectrum on a reflecting mirror with 
a specially designed mask to remove OH airglow lines. The reflected light, free 
of OH features, is contracted into white light and fed into a low-resolution 
spectrograph/camera to produce a near-infrared spectrum for very faint objects 
\citep{iwa01}. The spectrograph/camera module of this instrument, CISCO 
(Cooled Infrared Spectrograph and Camera for OHS), can be used 
separately as an imaging device at the Nasmyth focus \citep{mot02}.

\subsection{CIAO}
The CIAO is an instrument dedicated for diffraction-limited imaging of faint objects 
around a bright point source, such as circumstellar disks, circumstellar envelopes 
of evolved stars, and quasar host galaxies. It has a set of occulting masks as small 
as 0\farcs1 in diameter to allow imaging of structures very near the central source, 
utilizing the infrared adaptive optics system. The system provides both grism 
spectroscopic and polarimetric observational capabilities \citep{tam00}. The 
instrument has been commissioned, and the status of the instrument is described in 
\citet{tam00, tamu03, mura02}.

\subsection{COMICS}
The COMICS is an instrument housing 6 arrays of 320 ${\times}$ \rm 240 
Si:As IBC detectors for 
the 8--13 and 16--26 ${\mu m}$ \rm bands. It has an imaging capability of 0\farcs13 /pixel, covering a field of dimensions 42\arcsec $\times$ \rm 31\arcsec.  Long-slit 
spectroscopy covering up to 40\arcsec at 0\farcs165/pixel resolution is also available. 
The readout electronics newly developed for COMICS enabled this instrument to be the 
first mid infrared instrument with new-generation array offered for open use on 8m 
class telescopes \citep{kat00, okam03, sako02}.

\subsection{Adaptive Optics}
The Cassegrain adaptive optics (AO) is optimized for use in the \it{K} \rm band \citep{takm03a, takm03b}. IRCS and CIAO are designed to use this AO system. A bimorph deformable mirror 
driven by 36 electrodes is used to compensate for any wavefront aberration as measured 
by a curvature wavefront sensor with 36 avalanche photodiodes operated in the 
photon-counting mode. The wavefront sensing beam is divided and directed to the 36 
photodiodes through a custom-designed microlens array, so as to match the distribution 
of electrodes on the bimorph mirror. An oscillating membrane is used to generate the 
necessary offsets of the pupil image immediately in front and behind the detector 
array. The wavefront sensor is mounted directly on the instrument so as to minimize the
relative flexure between the wavefront sensor and the instrument. The unvignetted 
field of view of the AO system is 120\arcsec in diameter. The current AO system provides, 
with a guide star brighter than ${R_{\rm C}}$=13 mag, corrected images of 0\farcs065 in FWHM 
with a Strehl ratio of 0.4 in the \it{K} \rm band within a field of 20\arcsec in diameter. By 
inserting the retractable feed mirror in front of the Cassegrain focal plane, 
an observation can be switched easily between the ordinary mode and the AO mode without changing 
the focal position of the instruments or the \it{F} \rm ratio of the incident beam. The total 
bandwidth is about 100 Hz. Usage of a laser guide star is expected in 2005 to increase 
the sky coverage for AO \citep{hay00, haya02}.

\section{Operation Performances}
Although the Cassegrain first light was achieved at the end of 1998, commissioning 
all four foci and seven common-use scientific instruments took another two years. 
Limited open use of the Subaru Telescope was offered from 2000 December. For the first 
semester S00B, from 2000 December to 2001 March, only Suprime-Cam and CISCO were 
available. However, other instruments gradually became available, firstly with 
limited modes of operation, but eventually with all modes of observation planned 
for each instrument. Although the failure rates are gradually being reduced, further 
improvements are required to reach stable and efficient operation of the telescope 
and instruments. 

The telescope was closed for two months from 2001 August to September to address the 
problem of detached pads on the back of the primary mirror incorporated as part of 
the mechanical fuse system. The mirror was realuminized during this period and in 2003 
August.

\subsection{Active Mirror Support}
A block diagram of the active support system for the primary mirror is shown in 
figure~\ref{takaf1}.  The control system consists of two feedback loops. The main 
servo control of the support force loop is activated every 100 ms according to an 
open-loop force command issued by the real-time control computer. The computer updates 
the support force for the current elevation angle of the telescope based on a series 
of functions calibrated by mirror analysis, which is carried out when the focal 
configuration is changed. The mirror analysis is performed using a Shack--Hartmann 
camera installed at each focus. The displacement of 136 spot images of a bright star, 
produced by an array of microlenses with $14\times14$ subapertures, are measured to 
derive the surface error of the primary mirror. An exposure time of 60 s is adopted 
to achieve a wavefront measurement accuracy of 28 nm rms by averaging out atmospheric 
turbulence. The surface error is expanded by the eigenmodes of the primary mirror, 
and the coefficients of the first 32 eigenmodes are used to derive the optimum 
correction force to be added for each of the 261 actuators. Calibration of the active 
support system is established by performing such measurements at various 
orientations.

The actual dynamic performance of the mirror surface is measured by repeating the 
Shack--Hartmann wavefront measurements for 1 hour. Figure~\ref{takaf2} shows an 
example of such measurements carried out during 12:15--13:15 UT on 2001 September 27 at 
elevation angle $50^{\circ}-46^{\circ}$. The surface error on average is about 200 nm rms 
\citep{tako02}. The current rms surface error of the primary mirror is about 
$300-400$ nm, which corresponds to an image spread of $0\farcs15-0\farcs2$ at all elevation 
angles ($\sim 20^{\circ}$ to $85^{\circ}$). Under the best conditions, error of less 
than 100 nm has been achieved.

\begin{figure}
  \begin{center}
  \end{center}
  \caption{Dynamic performance of the residual surface error of the mirror as derived 
by contiguous Shack-Hartmann measurements.\label{takaf2}}
\end{figure}

There are three measures that can be taken to improve the residual surface error of 
the primary mirror: (1) The support force distribution can be fine tuned by taking the 
thermal-deformation effect of the primary mirror and the mirror cell into account 
in more detail. (2) A real-time feed back loop using light from an offset guide star 
directed to the Shack--Hartmann camera by a dichroic prism can be implemented. (3) 
A real-time fixed-point feedback (FFB) system 
can be introduced to reduce the wind load on the three fixed points \citep{kanz02}.

   \begin{figure}
   \begin{center}
   \end{center}
   \caption{Seeing size statistics. The diamonds, squares, and circles indicate the 
measurement time before 21:00, 21:00--26:00, and after 26:00 Hawaiian standard time, 
respectively. \label{seeingst}}
   \end{figure} 

  \begin{figure}
  \begin{center}
  \end{center}
  \caption{Monthly average seeing size during the 2000 June -- 2002 December period. \label{seeingmn}}
  \end{figure} 

\begin{figure}
  \begin{center}
  \end{center}
  \caption{Wind screen and wind sensors.\label{kanzf7}}
\end{figure}

\subsection{Image Quality}
The full width at half maximum (FWHM) seeing size measured by the CCD camera of the 
auto guider in the red band during focus checks has been recorded since 1999. 
Figure~\ref{seeingst} shows the seeing size statistics for the period 2000 May to 
2002 July. The median image size is $0\farcs6-0\farcs7$ FWHM in the ${R_{\rm C}}$ and ${I_{\rm C}}$ bands 
at all four foci. The best images obtained so far are 0\farcs2 FWHM without AO 
and 0\farcs065 with AO in the near-IR \it{K}\rm -band, and 0\farcs3 in the 
optical and mid-IR. This figure also shows that the seeing is generally slightly better 
at the later half of the night, probably due to better thermal equilibrium achieved 
with the ambient air.

Figure~\ref{seeingmn} shows the monthly average seeing size in arcsec as measured 
during the 2000 June -- 2002 December period. There is a repeated seasonal trend of seeing 
variation.  The best average seeing of 0\farcs5 is achieved during September, and May 
is the second-best month.

The temperature of the primary mirror and the enclosure are controlled during the 
day to a few degrees below the predicted night-time temperature so as to reduce the facility 
seeing. Approximately during a one-hour prior to the start of evening observation, all of the 
ventilators are opened to let air flow in the enclosure, thus assisting to reduce 
the temperature difference between the structure and the ambient air (cf. 
figure~\ref{kanzf7}). Figure~\ref{miyaf7} shows the measured temperature difference 
between the primary mirror and the ambient air. The details of the temperature control 
and the seeing statistics have been reported by \citet{miys02}.

\begin{figure}
  \begin{center}
  \end{center}
  \caption{Mirror temperature with respect to the ambient air 
temperature.\label{miyaf7}}
\end{figure}

   \begin{figure}[h]
   \begin{center}
   \end{center}
   \caption{Az track levels undulation as measured in 1997 January and in 
2002 February. The unevenness increased from 0.1 mm to 0.3 mm p-p, probably due to 
a persistent uneven load distribution.  A fine tuning of the rail to reduce this undulation is 
foreseen.\label{railcomp}}
   \end{figure} 

   \begin{figure*}
   \begin{center}
   \end{center}
   \caption{Control block diagram for the IR secondary mirror.\label{Chopping}}
   \end{figure*} 

\subsection{Pointing Performance}

The telescope can be pointed for elevation angles of  $89^{\circ} \ge El \ge 15^{\circ}$. However, 
to avoid any fast movement of the telescope near to the zenith, and large extinction at lower 
elevation angles, planning of an observation in the elevation angle range $85^{\circ} \ge El 
\ge 30^{\circ}$ is recommended. The blind pointing accuracy is better than 1\farcs0 
rms across most of the sky. The 30\arcsec field of view of the acquisition guide 
camera with a 0\farcs07 positioning accuracy is sufficient for all imaging observations.  
For spectroscopic observations, short acquisition images should be taken to ensure 
that the target remains on the center of the spectroscopic slit width. 

The parameters of the telescope pointing model equation are calibrated whenever the 
top unit is changed. The high-precision repeatability of the top unit exchange 
operation is proved by the stability in the offset values of the pointing model. Small 
systematic changes in some of the parameters were observed after aluminum re-coating 
and repairs on the fixed-points pads during the summer of 2001.

\subsection{Tracking Accuracy}
The Subaru Telescope runs on a circular azimuthal track rail. Six hydrostatic oil 
pads lift the 555 tonne telescope by about 50 $\mu$m, and six linear magnetic motors 
drive the telescope directly without any stick-slip. Due to the small discontinuity in 
the level of adjacent track rails, the tracking fails by as much as $\sim$2\arcsec 
when the pad passes these joints. Software correction using a lookup table was applied 
to address this problem \citep{tana98}, with the result that an open-loop tracking error 
of less than 0\farcs2 rms for 10 minutes has been achieved. Non-sidereal tracking 
of the telescope using a look-up table is also feasible. The tracking error has 
reportedly increased over the last two years, and is currently being investigated. 

Azimuthal track rails were installed in 1997 January with a peak-to-peak leveling 
error of 0.1 mm. In 2002 February, the vertical undulation of the azimuthal rails 
was found to have increased to 0.3 mm, attributed to uneven sinking of the non-shrink 
mortar underneath the azimuthal track rails. Figure~\ref{railcomp} shows the increase 
of the measured undulation of the track rail as a function of the telescope pointing 
\citep{usu02}. A newly implemented Az correction table has restored the open-loop 
tracking accuracy.

\subsection{Performances of Acquisition Guide System}

The measured guiding error, including the seeing effect, is less than 0\farcs1 rms with 
a guide star brighter than magnitude 16. The readout noise of the acquisition/guide 
camera at the Cassegrain and prime foci is $\sim$10 e$^{-1}$, while that at the Nasmyth 
foci is as high as 100 e$^{-1}$. The acquisition time has been shortened to less than 
1--2s.

\subsection{Infrared Secondary Mirror}

The tip-tilt and chopping infrared secondary mirror \citep{itoh98} is made of 
light-weight ULE glass (185 kg) and has a silver coating.  Figure~\ref{Chopping} shows 
the control block diagram for the IR secondary mirror. The driving mechanism consists 
of 6 electro-magnetic actuators, 15 bellows for passive support, 3 electric 
capacitance sensors, and a reaction force compensator. The maximum chopping frequency 
currently available is 3~Hz with a duty cycle of 80\% and a position error of $< 
0\farcs1$ rms. The chopping amplitude is $60\arcsec$, and the sampling rate and detection rate 
of the sensors are both 1 ms.

The rms residual guiding error with the tip-tilt secondary mirror was measured to 
be 0\farcs028 using the fast guiding system with a guide star of magnitude 11--14.

   \begin{figure}[h]
   \begin{center}
   \end{center}
   \caption{Binary star image taken using the FOCAS instrument on 2000 August 9. 
The upper star is a red star, and the lower is a blue star. The image of the blue 
star is clearly elongated when the ADC is not at correct rotation angle. The seeing 
was about $\sim0\farcs6$. \label{ADC}}
   \end{figure} 

   \begin{figure}[h]
   \begin{center}
   \end{center}
   \caption{Reflectivity and surface roughness degradation of the primary mirror 
measured at 670 nm. \label{M1_monitor}}
   \end{figure} 

\subsection{Atmospheric Dispersion Correction and Image Derotation}
The atmospheric dispersion corrector (ADC) units, comprising a counter-rotating pair 
of direct-vision prisms with matching silicon oil, are installed at the Cassegrain 
and Nasmyth foci to compensate for atmospheric dispersion down to an elevation angle 
of $30^{\circ}$. The ADC for the prime focus is based on different optics \citep{nar94} 
and is assembled in the top unit module for the Suprime-Cam.  Figure~\ref{ADC} shows 
test images for a pair of blue and red stars taken using the FOCAS \citep{kas02} at 
an elevation of $55^{\circ}$ with the ADC at various settings. The image of the blue star 
is clearly elongated when the ADC is not at the correct rotation angle.

The instrument rotators for field derotation are installed at the prime focus and 
Cassegrain focus, while three-mirror image derotators are equipped at Nasmyth-Opt and Nasmyth-IR foci. The derotation error at the prime focus and 
Cassegrain focus is practically negligible. The derotation error at the Nasmyth foci 
is less than $< 0.^{\circ}02 \rm hr^{-1}$.

\subsection{Reflectivity of the Primary Mirror}
The latest realuminization of the 8.3 m primary mirror was conducted in 2003 August. 
It was the fourth such maintenance operation conducted since the arrival of the mirror 
at the summit in 1998. As a maintenance routine, the mirror is cleaned with CO$_2$ dry 
ice every 2 to 3 weeks under lower (50\%) humidity conditions using an in situ cleaning 
device consisting of 4 deployable arms mounted at the lower edge of the center section. 
These arms are fitted with many nozzles to supply dry ice to the mirror surface, and 
are swept over the entire mirror surface to remove dust accumulated on the mirror 
surface \citep{tor98}. Figure~\ref{M1_monitor} shows how the reflectivity of the 
primary mirror decreases over time. At optical wavelengths ($\lambda = 670$ nm), the 
reflectivity is kept better than 82-83\% and the surface roughness is $70-80 {\rm 
\AA}$ as measured using a scatter meter ($\mu$Scan, TMA Technologies, Inc.).

   \begin{figure}[h]
   \begin{center}
   \end{center}
   \caption{CIAX cart for Cassegrain instrument exchange operation dismounts an 
instrument from the Cassegrain focus and brings it to one of the four Cassegrain 
instrument standby ports. CIAX then takes the next instrument from the standby port 
and mounts it back to the Cassegrain focus.\label{ciax}}
   \end{figure} 

   \begin{figure}[h]
   \begin{center}
   \end{center}
   \caption{Two of the four Cassegrain instruments standby ports.  FOCAS is mounted 
on one of the ports, but the other port for IRCS is unoccupied in this picture. 
\label{stdbypr1}}
   \end{figure} 

\subsection{Instrument Exchange}

Exchange of the Cassegrain instruments is achieved effectively and smoothly with using 
the Cassegrain Instrument Automatic eXchanging system \citep{usu00}. 
Figure~\ref{ciax} shows the CIAX cart carrying one of the Cassegrain instruments, 
CIAO, during the exchange operation. This can be performed during the day, and takes 
about two hours. There are four stand-by platform ports on the Cassegrain floor where 
the remaining three instruments can be kept alive with an electric and cryogenic power 
supply as well as connection to the network system. Figure~\ref{stdbypr1} shows two 
such stations where one of the stations is occupied by FOCAS. 

Exchanging the top unit requires 3--6 hr of work during the day. Linking and resetting 
the secondary mirror requires an additional 30 min. Inserting the tertiary mirror 
takes only 2 min. Rotation of the enclosure takes less than 6 min. Although 
the level of operation of the telescope and instruments have become smoother and more 
stable over the last two years, there are still, naturally, many areas where 
improvements can be made to increase operational efficiency.

During the 28 months between 2000 May and 2002 August, the telescope was operational 
for 26 months. The top unit was exchanged 56 times and the instruments were changed 
89 times, corresponding to roughly 2.2 top unit exchange operations and 3.4 instrument 
exchange operations per month.  

   \begin{figure}
   \begin{center}
   \end{center}
   \caption{Increase of science exposure time fraction in Suprime-Cam observations. 
``Readout + Calib" is the time fraction for calibration exposures and CCD readouts 
for both science frames and calibration frames. ``Others" include time for telescope 
pointing, setting up for a guide star, filter exchange, and other preparatory works.  
It must be noted that telescope idling time waiting for expectant improvement of poor 
weather condition is also included in ``Other". \label{noumaru}}
   \end{figure} 

\subsection{Efficiency of Operation}
Software revisions were made to reduce the time required for pointing
the telescope by improving the control parameters and algorithm. Also 
made were modifications of the drive mechanisms and the procedure to 
speed up the top unit exchange and Cassegrain instrument exchanges. 
Together with these efforts to improve the operation efficiency of the 
telescope, separate efforts on instruments have been made during the last
few years.

Figure~\ref{noumaru} shows the fraction of science exposure time for
Suprime-Cam observations during the time when the telescope dome was
open \citep{nou02}. The minimum overhead between successive CCD exposures
used to be about 240 s for the Suprime-Cam, which includes CCD wipe out and the signal
readout. The overhead became shortened to 120s by replacing 
SITe CCDs to MIT/LL CCDs in 2001 April, and was reduced eventually 
to 60s by virtue of newly developed readout electronics; Messia-V
(Komiyama et al. 2003). One can read in this figure the 
tendency of improved shutter open time fraction in accordance with 
these improvements.  The actual fraction of time spent on target 
exposure measured versus the time with the telescope pointed to the 
target was of course even larger and is now about 75\%, depending
on the filters and the observed objects.

   \begin{figure}
   \begin{center}
   \end{center}
   \caption{Nights assigned for open-use programs in five categories of science 
observation. \label{scieuse}}
   \end{figure} 

   \begin{figure}
   \begin{center}
   \end{center}
   \caption{Numbers of assigned nights for open-use programs requesting each of the 
open-use instrument.  Because the Adaptive Optics is usually used either with IRCS or with 
CIAO, those nights are doubly counted on each instrument.\label{instuse}}
   \end{figure} 

\begin{longtable}{lllll}
  \caption{Statistics on open-use application.}\label{proposal}
  \hline\hline
  Semester & Submitted & Adopted & Requested & Assigned \\
    & proposals & proposals & nights & nights \\
  \hline
  S00B &  114 & 26& 223 & 36 \\
  S01A &  105 & 27& 204 & 36 \\
  S01B &  160 & 29 & 337 & 47 \\
  S02A &  186 & 37 & 410 & 69 \\
  S02B &  193 & 38 & 448 & 74 \\
  S03A &  195 & 40 & 440 & 76 \\
  S03B &  196 & 45 & 553 & 94.5 \\
  \hline
\endlastfoot
\end{longtable}



\subsection{Open-Use Statistics}
The Subaru Telescope was made available for open use in 2000 December (S00B semester), 
with only Suprime-Cam and CISCO offered to the community. Now, all seven open-use 
instruments and the Cassegrain adaptive optics system are available for open use. 
Some of the instruments, however, are offered with some operational restrictions. 
All applications for open-use time are reviewed by referees nominated by the Time 
Allocation Committee (TAC), and the TAC prepares a list of successful proposals.  The 
competition rate over the last seven semesters has on average been 5 times for proposals 
and 6 times for nights. Table~\ref{proposal} gives the statistics for open-use 
applications. In order to increase the opportunity for potential users to obtain 
observing experience using the Subaru Telescope, TAC decided to accept only short 
programs requiring no more than 3 nights for the first three semesters. However, from 
semester S02A, larger scale programs with a maximum of 10 nights have been considered 
to derive science outcomes on systematic studies. Figure~\ref{scieuse} shows the 
number of assigned open-use nights in five major fields for semesters S00B--S03B, 
and figure~\ref{instuse} shows the number of assigned open-use nights for individual 
instruments for semesters S00B--S03B.

In addition to these open-use programs, two major coordinated projects, the Subaru 
Deep Field Survey (SDF), the Subaru--XMM Deep Field Survey (SXDS), and the Subaru Disk and Planet Search (SDPS) have been organized 
and are conducted by the observatory staff, for which up to 33 nights per year in 
total have been dedicated until S03B. 

Science outputs from Subaru Telescope are published in many papers.  Short reviews 
on some of the scientific achievements are made \citep{iye02}.

\section{Future Plan}
Second-generation instruments, FMOS and MOIRCS, are currently under construction. FMOS 
is a fiber multi-object spectrograph for the \it{J} \rm and \it{H} \rm bands to be mounted at the prime 
focus for spectroscopic observation of up to 400 objects in a 30' field 
\citep{mai00}. Two sets of Echidna, each having 200 fiber head positioners installed 
in the focal plane unit, pick up objects and relay the sampled light to two infrared 
spectrographs. Construction is under way in collaboration with Kyoto University, the 
UK, and the Anglo-Australian Observatory (AAO), and is to be commissioned in 2005. 
MOIRCS is a fully cryogenic double beam multi-object infrared camera and spectrograph 
that is presently being constructed as a joint project of NAOJ and Tohoku University.  
First light for the camera section is expected in 2004.  

A five year project to upgrade the currently offered AO system was approved and a 
special Grant-in-Aid for the Promotion of Science has been awarded by the Ministry 
of Education, Science, Culture, Sports and Technology. The new system will have 
five-times more control elements, increasing the Strehl ratio of the point spread function. 
Development of a laser guide star system is another key feature of this project, and 
will increase the sky coverage of the AO system.

To improve operational efficiency and optimize the observation programs according 
to the weather conditions, service observation will be offered on a limited scale 
from S03A, and flexible scheduling is under investigation. Use of the optical 
Cassegrain secondary mirror for the Nasmyth focus by applying active deformation of 
the primary mirror to remove any spherical aberration is also currently under 
investigation as a means to reduce the need to exchange the secondary mirrors.  
Like every other telescope, there are many on-going improvements on Subaru Telescope 
and its instrumentation, and some of the descriptions of the present paper will definitely 
become out of date soon.  The most recent information on Subaru Telescope can 
be checked at the following web site: http://subarutelescope.org/index.html.  Readers are encouraged to obtain the most recent information from this site and/or direct 
contact to the Subaru Telescope staff.

\section{Acknowledgements}
The Subaru Telescope project has been a tremendous undertaking that has become a 
working reality through the dedication of the many people involved. Besides the 
astronomers, engineers, and administrative staff of NAOJ, there are many scientists 
from other universities, administrative staff from the Japanese government, and 
engineers and managing staff from contracted industries, who played valuable roles 
in various parts and stages of the project. All of the people who contributed to the 
construction and operation of the Subaru Telescope are greatly appreciated. 

The project team owes much among others to Drs. D. Hall, R. MacLaren, and A. Tokunaga 
of the Institute of Astronomy, University of Hawaii, for their determined promotion 
of Subaru Telescope project, Dr. S. Okamura of University of Tokyo, Drs. H. Ohtani and 
T. Maihara of Kyoto University for their essential contribution to support the project.

Special thanks are extended to Messrs. N. Itoh, I. Mikami, and O. Sakakibara of 
Mitsubishi Electric Corporation for their originality and persistence to complete 
the telescope, Mr. K. Kawarai of Fijitsu Limited, Mr. J. Kawai of Fujitsu America, 
Mr. R. Smith of Corning Incorporated, Dr. S. Smith and Mr. M. Young of Contraves 
Brashear Corporation for their essential contributions to the success of the project. 


\end{document}